\documentclass[11pt]{article}
\usepackage[left=1.25in,top=1in,right=1.25in,bottom=1in]{geometry}
\usepackage{amsmath}
\usepackage{amssymb}
\usepackage{amsbsy}
\usepackage{amsthm}
\usepackage{epsfig}
\usepackage[round]{natbib}
\usepackage{clrscode}
\usepackage{setspace}
\usepackage{multirow}
\usepackage{color}
\usepackage{url}
\usepackage{booktabs}

\usepackage{rotating}

\newcommand{\N}{\mbox{N}}

\newcommand{\bbeta}{\boldsymbol{\beta}}

\newcommand{\nc}{\newcommand}

\nc{\norm}{\mathcal{N}}
\nc{\gig}{\mathcal{G}}
\nc{\pol}{\mathcal{P}}
\nc{\gh}{\mathcal{GH}}
\nc{\ig}{\mathcal{IG}}
\nc{\bX}{{ X}}
\nc{\lhat}[1][i]{\hat\lambda_{#1}}
\nc{\what}[1][j]{\hat\omega_{#1}}
\nc{\bone}{\vec{1}}
\nc{\Li}{\hat\Lambda}
\nc{\Oi}{\hat\Omega}
\nc{\diag}[1]{\text{diag}\left(#1\right)}
\nc{\Siginv}{\Sigma^{-1}}
\nc{\Ominv}{\Omega^{-1}}
\nc{\ms}{\mathcal{MS}}

\newtheorem{theorem}{Theorem}
\newtheorem{proposition}[theorem]{Proposition}

\title{Data augmentation for non-Gaussian regression models using variance-mean mixtures}

\author{
\textsc{Nicholas G.~Polson} \\
\small{\textit{Booth School of Business}} \\ \small{\textit{University of Chicago}} \\
\\
\textsc{James G.~Scott} \\
\small{\textit{University of Texas at Austin}} \\
}

\begin{document}
\begin{spacing}{1.1}

\maketitle

\begin{abstract}
We use the theory of normal variance-mean mixtures to derive a data-augmentation scheme for a class of common regularization problems.  This generalizes existing theory on normal variance mixtures for priors in regression and classification.  It also allows variants of the expectation-maximization algorithm to be brought to bear on a wider range of models than previously appreciated.  We demonstrate the method on several examples, including sparse quantile regression and binary logistic regression.  We also show that quasi-Newton acceleration can substantially improve the speed of the algorithm without compromising its robustness.
\vspace{1pc}

\end{abstract}
\section{Introduction}

\subsection{Regularized regression and classification}

Many problems in regularized estimation involve an objective function of the form
\begin{equation}
\label{eqn:generalobjectivefunction}
L(\beta) =   \sum_{i=1}^n f( y_i , x_i^T \beta \mid \sigma) 
 + \sum_{j=1}^p g (\beta_j \mid \tau)  \, .
\end{equation}
Here $y_i$ is a response, which may be continuous, binary, or multinomial; $x_i$ is a known $p$-vector of predictors; $\beta = ( \beta_1 , \ldots , \beta_p )$ is an unknown vector of coefficients; $f$ and $g$ are the negative log likelihood and penalty function, respectively; and $\sigma$ and $\tau$ are scale parameters, for now assumed fixed.  We may phrase the problem as one of minimizing $L(\bbeta)$, or equivalently maximizing the unnormalized posterior density $\exp\{-L(\bbeta)\}$, interpreting the penalty as a negative log prior.

In this paper, we unify many disparate problems of this form into a single class of normal variance-mean mixtures.  This has two practical consequences.  First, it facilitates the use of penalty functions in models with rich structure, such as hierarchical non-Gaussian models, discrete mixtures of generalized linear models, and missing-data problems.  

Second, it can be used to derive simple expectation-maximization algorithms for non-Gaussian models. This leads to a stable, unified approach to estimation in many problems, including quantile regression, logistic regression, negative-binomial models for count outcomes, support-vector machines, and robust regression.

The key result is Proposition \ref{thm:reptheorem1}, which describes a simple relationship between the derivatives of $f$ and $g$ and the conditional sufficient statistics that arise in our expectation-maximization algorithm.  The expected values of these conditional sufficient statistics can usually be calculated in closed form, even if the full conditional distribution of the latent variables is unknown or intractable.    Proposition \ref{thm:reptheorem3} provides the correponding result for the posterior mean estimator, generalizing the results of \citet{masreliez:1975} and \citet{pericchi:smith:1992}.

One known disadvantage of expectation--maximization algorithms is that they may converge slowly, particularly if the fraction of missing information is large \citep{meng:vandyk:1997}.  Our method, in its basic form, is no exception.  But we find that substantial gains in speed are possible using quasi-Newton acceleration \citep{lange:1995}.  The resulting approach combines the best features of the expectation--maximization and Newton--Raphson algorithms: robustness and guaranteed ascent when far from the solution, and super-linear convergence when close to it.

\subsection{Relationship with previous work}

Our work is motivated by recent Bayesian research on sparsity-inducing priors in linear regression, where $f = \Vert y - X \bbeta \Vert^2$ is the negative Gaussian log likelihood, and $g$ corresponds to a normal variance-mixture prior \citep{andrews:mallows:1974}.  Examples of work in this area include the lasso \citep{tibshirani1996,park:casella:2008,hans:2008}; the bridge estimator \citep{west:1987,knight:fu:1998,huang:horowitz:ma:2008}; the relevance vector machine of \citet{tipping:2001}; the normal/Jeffreys prior of \citet{figueiredo:2003} and \citet{bae:mallick:2004}; the normal/exponential-gamma model of \citet{griffin:brown:2005}; the normal/gamma and normal/inverse-Gaussian models \citep{caron:doucet:2008, griffin:brown:2010}; the horseshoe prior of \citet{Carvalho:Polson:Scott:2008a}; the hypergeometric inverted--beta model of \citet{polson:scott:2009a}; and the double-Pareto model of \citet{dunson:armagan:lee:2010}.  See \citet{Polson:Scott:2010a} for a review.

We generalize this work by representing both the likelihood and prior as variance-mean mixtures of Gaussians.  This data-augmentation approach relies upon the following decomposition:
\begin{eqnarray*}
p( \beta \mid \tau , \sigma, y ) \propto  e^{-L(\bbeta)}  &\propto& \exp \left\{ - \sum_{i=1}^n f ( y_i , x_i^T \beta \mid \sigma ) 
 - \sum_{j=1}^p g ( \beta_j \mid \tau  ) \right\} \\
 & \propto& \left\{ \prod_{i=1}^n p( z_i \mid \bbeta, \sigma ) \right\} \left\{  \prod_{j=1}^p p( \beta_j \mid  \tau ) \right\} \nonumber = p( z \mid \beta, \sigma) \ p( \beta \mid \tau ) \nonumber \, ,
\end{eqnarray*}
where the working response $z_i$ is equal to $y_i - x_i^T \beta $ for Gaussian regression, or $y_i x_i^T \beta $ for binary classification using logistic regression or support-vector machines.  Here $y_i$ is coded as $\pm 1$.  Both $\sigma$ and $\tau$ are hyperparameters; they are typically estimated jointly with $\bbeta$, although they may also be specified by the user or chosen by cross-validation.  In some cases, most notably in logistic regression, the likelihood is free of hyperparameters, in which case $\sigma$ does not appear in the model.  Previous studies \citep[e.g.][]{polson:stevescott:2011,gramacy:polson:2012} have presented similar results for specific models, including support-vector machines and the so-called powered-logit likelihood.  Our paper characterizes the general case.

One thing we do not do is to study the formal statistical properties of the resulting estimators, such as consistency as $p$ and $n$ diverge.  For this we refer the reader to \citet{fan:li:2001} and \citet{griffin:brown:2005}, who discuss this issue from classical and Bayesian viewpoints, respectively.

\section{\label{sec:varmixnorm}Normal variance-mean mixtures}

\begin{table}[t]
\begin{center}
\begin{footnotesize}
\caption{Variance-mean mixture representations for many common loss functions.  Recall that $z_i = y_i - x_i^T \beta $ for regression, or $z_i = y_i x_i^T \beta $ for binary classification. \label{tab:commonpenalties}}
\vspace{0.5\baselineskip}
\begin{tabular}{l l l l l}
Error/loss function & $f(z_i \mid \bbeta, \sigma)$ & $\kappa_z$ & $\mu_z$ & $p(\omega_i)$ \\
Squared-error & $z_i^2/\sigma^2 $ & 0 & 0 & $\omega_i = 1$ \\ 
Absolute-error & $ |z_i/\sigma| $ & 0 & 0 & Exponential \\ 
Check loss & $|z_i|  + (2q-1)z_i $ & $1-2q$ & 0 & Generalized inverse Gaussian \\ 
Support vector machines & $\max (1-z_i, 0)$ & $1$ & 1 & Generalized inverse Gaussian \\ 
Logistic & $\log (1+e^{z_i})$ & $1/2$ & 0 & Polya
\end{tabular}
\end{footnotesize}
\end{center}
\end{table}

There are two key steps in our approach.  First, we use variance-mean mixtures, rather than just variance mixtures.  Second, we interweave two different mixture representations, one for the likelihood and one for the prior.  The introduction of latent variables $\{\omega_i\}$ and $\{\lambda_j\}$ in Equations (\ref{eqn:likelihoodmixture}) and (\ref{eqn:priormixture}), below, reduces $\exp\{-L(\bbeta)\}$ to a Gaussian linear model with heteroscedastic errors:
\begin{align}
p( z_i \mid \beta, \sigma ) & = 
\int_0^\infty \phi ( z_i \mid \mu_z + \kappa_z \omega_i ^{-1}, \sigma^2 \omega_i ^{-1}) 
\ d P(\omega_i) \label{eqn:likelihoodmixture} \, ,\\
p( \beta_j \mid \tau ) & = 
\int_0^\infty \phi ( \beta_j \mid \mu_{\beta} + \kappa_{\beta} \lambda_j ^{-1}, \tau^2\lambda_j ^{-1}) 
\ d P(\lambda_j) \label{eqn:priormixture} \, ,
\end{align}
where $\phi(a \mid m, v)$ is the normal density, evaluated at $a$, for mean $m$ and variance $v$.

By marginalizing over these latent variables with respect to different fixed combinations of ($\mu_z, \kappa_z, \mu_{\beta}, \kappa_{\beta})$ and the mixing measures $P(\lambda_j)$ and $P(\omega_i)$, it is possible to generate many commonly used objective functions that have not been widely recognized as Gaussian mixtures.    Table \ref{tab:commonpenalties} lists several common likelihoods in this class, along with the corresponding fixed choices for $(\kappa_z, \mu_z)$ and the mixing distribution $P(\omega_i)$.

A useful fact is that one may avoid dealing directly with conditional distributions for these latent variables.  To find the posterior mode, it is sufficient to use Proposition \ref{thm:reptheorem1} to calculate moments of these distributions.  These moments in turn depend only upon the derivatives of $f$ and $g$, along with the hyperparameters.

We focus on two choices of the mixing measure: the generalized inverse-Gaussian distribution and the Polya distribution, the latter being an infinite sum of exponential random variables \citep{bn:kent:sorensen:1982}.  These choices lead to the hyperbolic and $Z$ distributions, respectively, for the variance-mean mixture.  The two key integral identities are
\begin{align}
 \frac{\alpha^2 - \kappa^2}{2 \alpha} e^{-\alpha | \theta - \mu | + \kappa ( \theta - \mu ) }
 & = \int_0^\infty \phi \left (\theta \mid   \mu + \kappa v ,  v \right ) 
 \ p_{\gig} \big\{ v \mid 1 , 0 , ( \alpha^2 - \kappa^2)^{1/2} \big\}  \ d v \, , \; \mbox{and} \nonumber \\
\frac{1}{B( \alpha , \kappa )} 
  \frac{ e^{ \alpha (\theta- \mu) } }
  { ( 1 + e^{\theta-\mu} )^{ 2 (\alpha - \kappa) } }
 & = \int_0^\infty \phi \left (\theta \mid \mu + \kappa v , v \right )
\ p_{\pol} ( v \mid \alpha , \alpha - 2 \kappa) \ d v \, , \label{eqn:logitmixture}
\end{align}
where $p_{\gig}$ and $ p_{\pol} $ are the density functions of the generalized inverse-Gaussian and Polya distributions, respectively.  For details, see Appendix \ref{app:mixturetheory}.   We use $\theta$ to denote a dummy argument that could involve either data or parameters, and $v$ to denote a latent variance.  All other terms are pre-specified in order to represent a particular density or function.  These two expressions lead, by an application of the Fatou--Lebesgue theorem, to three further identities for the improper limiting cases of the two densities above:
\begin{align*}
a^{-1} \exp \left\{ - 2 c^{-1} \max ( a\theta , 0 ) \right\} &= \int_0^\infty \phi ( \theta \mid - a v , c v ) \ d v  \, , 
\\
 c^{-1} \exp \left\{ - 2 c^{-1} \rho_q (  \theta ) \right\}  &= \int_0^\infty \phi ( \theta \mid  v - 2\tau v , c v ) e^{ -  2 \tau ( 1 - \tau ) v } \ d v \, ,  
 \\
\left( 1 + \exp\{ \theta-\mu \} \right)^{-1} & =
\int_0^\infty \phi(\theta \mid \mu - v/2, v) \
p_{\pol}( v \mid 0,1 ) \ dv   \, , 
\end{align*}
where $ \rho_q ( \theta ) = | \theta |/2 + \left ( q - 1/2 \right ) \theta $ is the check-loss function.  The first leads to the pseudo-likelihood for support-vector machines, the second to quantile and lasso regression, and the third to logistic and negative binomial regression under their canonical links.   The function $p_{\pol}( v \mid 0, 1 )  $ is an improper density corresponding to a sum of exponential random variables.  In the case of a multinomial model, the conditional posterior for a single category, given the parameters for other categories, is a binary logistic model.

\section{\label{sec:em-ecme}An expectation-maximization algorithm}

\subsection{Overview of approach}

By exploiting conditional Gaussianity, models of the form (\ref{eqn:likelihoodmixture})--(\ref{eqn:priormixture}) can be fit using an expectation-maximization algorithm \citep{dempster:laird:rubin:1977}.  In the expectation step, one computes the expected value of the log posterior, given the current estimate $\bbeta^{(g)}$:
$$
C(\beta\mid\beta^{(g)})  
= \int \log p(\beta \mid \omega, \lambda, \tau, \sigma, z) p(\omega, \lambda\mid \beta^{(g)}, \tau , z)  \ d\omega \ d\lambda \, .
$$
Then in the maximization step, one maximizes the complete-data posterior as a function of $\bbeta$.  The same representation can also be used to derive Markov-chain sampling algorithms, although we do not explore this aspect of the approach. 

The expectation-maximization algorithm has several advantages here compared to other iterative methods.  It is highly robust to the choice of initial value, has no user-specified tuning constants, and leads to a sequence of estimates $\{\beta^{(1)}, \beta^{(2)}, \dots \}$
that monotonically increase the observed-data log posterior density.  The disadvantage is the potential for slow convergence.  For the models we consider, this disadvantage is real, but can be avoided using quasi-Newton acceleration.

The complete-data log posterior can be represented in two ways.  First, we have
\begin{align}
  \log p( \beta \mid \omega, \lambda, \tau, \sigma, z )
   = c_0(\omega, \lambda, \tau, \sigma, z)
  & -\frac{1}{2 \sigma^2}\sum_{i=1}^n
 \omega_i \left( z_i - \mu_z - \kappa_y \omega_i^{-1}  \right)^2 \nonumber \\
    & - \frac{1}{2  \tau^2} \sum_{j=1}^p \lambda_j (\beta_j - \mu_{\beta} - \kappa_{\beta} \lambda_j^{-1}   )^2 \label{eqn:compdatalogpost}
\end{align}
for some constant $c_0$, recalling that $z_i = y_i - x_i^T \beta$ for regression or $z_i = y_i x_i^T \beta$ for classification.  Factorizing this further as a function of $\bbeta$ yields a second representation:
\begin{align}
  \log p( \beta \mid \omega, \lambda, \tau, \sigma, z )
   = c_1(\omega, \lambda, \tau, \sigma, z)
  & -\frac{1}{2\sigma^2} \sum_{i=1}^n
  \omega_i \left ( z_i - \mu_z \right )^2 
  +  \kappa_y  \sum_{i=1}^n \left ( z_i - \mu_z \right ) \nonumber \\
    & - \frac{1}{2 \tau^2} \sum_{j=1}^k \lambda_j (\beta_j - \mu_{\beta} )^2 + \kappa_{\beta} \sum_{j=1}^p 
( \beta_j - \mu_\beta ) \, \label{compdatalogpost2}
\end{align}
for some constant $c_1$.  We now derive the expectation and maximization steps, along with the complete-data sufficient statistics.

\subsection{The expectation step}

From (\ref{compdatalogpost2}), observe that the complete-data objective function is linear in both $\omega_i$ and $\lambda_j$.  Therefore, in the expectation step, we replace $\lambda_j$ and $\omega_i$ with their conditional expectations $\hat{\lambda}_j^{(g) }$ and $\hat \omega_i^{(g)}$, given the data and the current $\beta^{(g)}$.    The following result provides expressions for these conditional moments under any model where both the likelihood and prior can be represented by normal variance-mean mixtures.

\begin{proposition}
\label{thm:reptheorem1}
Suppose that the objective function $L(\bbeta)$ in (\ref{eqn:generalobjectivefunction}) can be represented by a hierarchical variance-mean Gaussian mixture, as in Equations (\ref{eqn:likelihoodmixture}) and (\ref{eqn:priormixture}).  Then the conditional moments $\hat \lambda_j = E \left ( \lambda_j \mid \beta, \tau, z   \right )  $ and $  \hat\omega_i = E \left ( \omega_i \mid \sigma, z_i \right ) $ are given by
$$
       ( \beta_j - \mu_{\beta} ) \hat\lambda_j = \kappa_{\beta} + \tau^2  \
   g'( \beta_j \mid \tau ) \; , \quad
      ( z_i - \mu_z ) \hat\omega_i  = \kappa_z + \sigma^2 \
    f'( z_i \mid \beta, \sigma)     \, ,
$$
where $f'$ and $g'$ are the derivatives of $f$ and $g$ from (\ref{eqn:generalobjectivefunction}).
\end{proposition}

This characterizes the required moments in terms of the likelihood $f(z_i) = -\log p(z_i \mid \beta, \sigma)$ and penalty function $g(\beta_j) = -\log p(\beta_j \mid \tau)$.  One caveat is that when $  \beta_j - \mu_{\beta} \approx 0 $, the conditional moment for $\lambda_j$ in the expectation step may be numerically infinite, and care must be taken.   Indeed, infinite values for $\lambda_j$ will arise naturally under certain sparsity-inducing choices of $g$, and indicate that the algorithm has found a sparse solution.  One way to handle the resulting problem of numerical infinities is to start the algorithm from a value where $(\beta - \mu_{\beta})$ has no zeros, and to remove $\beta_j$ from the model when it gets too close to its prior mean.  This conveys the added benefit of hastening the matrix computations in the maximization step.  Although we have found this approach to work well in practice, it has the disadvantage that a variable cannot re-enter the model once it has been deleted.  Therefore, once a putative zero has been found, we propose small perturbations in each component of $\bbeta$ to assess whether any variables should re-enter the model.  Our method does not sidestep the problem of optimization over a combinatorially large space.  In particular, it does not guarantee convergence to a global maximum if the penalty function is not convex, in which case restarts from different initial values will be necessary to check for the presence of local modes.

\subsection{The maximization step}

Returning to (\ref{eqn:compdatalogpost}), the maximization step involves computing the posterior mode under a heteroscedastic Gaussian error model and a conditionally Gaussian prior for $\bbeta$.
\begin{proposition}
\label{thm:reptheorem2}
Suppose that the objective function $L(\bbeta)$ in (\ref{eqn:generalobjectivefunction}) can be represented by variance-mean Gaussian mixtures, as in (\ref{eqn:likelihoodmixture})-(\ref{eqn:priormixture}).  Then given estimates $\hat\omega_i$ and $\hat\lambda_j$, we have the following expressions for the conditional maximum of $\beta$, where $\omega = (\omega_1, \ldots, \omega_n)$ and $\lambda = (\lambda_1, \ldots, \lambda_p)$ are vectors, and where $\Omega  = \mbox{diag}(\omega_1, \ldots, \omega_n )$ and $\Lambda = \mbox{diag}(\lambda_1, \ldots, \lambda_p)$.
\begin{enumerate}
\item In a regression problem, 
$$
\hat \beta = \big( \tau^{-2} \Li + \bX^T\Oi\bX \big) ^{-1}  (y^{\star} + b^{\star}) \, , \; 
y^{\star} = \bX^T \left( \Oi y - \mu_z \omega - \kappa_z \bone \right) \, , \;
b^{\star} = \tau^{-2} ( \mu_\beta \lambda + \kappa_{\beta} \bone_n) \, ,
$$
where $\bone_n$ indicates a vector of ones.
\item In a binary classification problem where $y_i = \pm 1$ and $\bX_{\star}$ has rows $x^{\star}_i = y_i x_i$,
$$
\hat \beta= \left( \tau^{-2} \Li + \bX_{\star}^T\Oi \bX_{\star} \right)^{-1} \bX_{\star}^T 
  \left( \mu_{z} \hat\omega + \kappa_{z} \bone \right) \, .
$$
\end{enumerate}
\end{proposition}

One may also use a series of conditional maximization steps for the regression coefficients $\bbeta$ and for hyperparameters such as $\sigma$ and $\tau$.  These latter steps exploit standard results on variance components in linear models, which can be found in, for example, \citet{gelman:2006}.

\subsection{Quasi-Newton acceleration}

There are many strategies for speeding up the expectation--maximization algorithm.  A very simple approach that requires no extra analytical work, at least in our case, is to use quasi-Newton acceleration.  The idea is to decompose the observed-data likelihood as $L(\bbeta) = C(\bbeta) - R(\bbeta)$, where $C(\bbeta)$ is the complete-data contribution and $H(\bbeta)$ is an unknown remainder term.  This leads to a corresponding decomposition for the Hessian of $L(\bbeta)$: $-\nabla^2 L(\bbeta) = -\nabla^2 C(\bbeta) + \nabla^2 R(\bbeta)$.  In models of the form (\ref{eqn:likelihoodmixture})--(\ref{eqn:priormixture}), $-\nabla^2 C(\bbeta)$ is simply the inverse covariance matrix of the conditional normal distribution for $\bbeta$, already given.  The Hessian of the remainder, $\nabla^2 R(\bbeta)$, can be iteratively approximated using a series of numerically inexpensive rank-one updates.  The approximate Hessian $\tilde{H} =  \nabla^2 L(\bbeta) - \nabla^2 R(\bbeta)$ is then used in a Newton-like step to yield the next iterate, along with the next update to $\nabla^2 R(\bbeta)$.

We omit the updating formula itself, along with the strategy by which one ensures that $-\nabla^2 L(\bbeta)$ is always positive definite and that each iteration monotonically increases the posterior density.  A full explanation for the general case can be found in \citet{lange:1995}.  Below, however, we show that quasi-Newton acceleration can offer a dramatic speed-up for variance-mean mixtures, compared to a na\"ive implementation of expectation--maximization.

\section{Examples}
\label{sec:examples}

\subsection{Logistic regression}

Suppose we wish to fit a logistic regression where
$$
\hat{\beta} = \arg \min_{\beta \in \mathbb{R}^p} \left[  \sum_{i=1}^n \log\{1 + \exp(-y_i x_i^T \beta) \}
 + \sum_{j=1}^p g(\beta_j \mid \tau)  \right] \, ,
$$
assuming that the outcomes $y_i$ are coded as $\pm1$, and that $\tau$ is fixed.  To represent the binary logistic likelihood as a mixture, let $\omega^{-1}$ have a Polya distribution with $ \alpha =1 , \kappa =  1/2 $.
Proposition \ref{thm:reptheorem1} gives the relevant conditional moment as
$$
\hat\omega_i  = \frac{1}{z_i} \left( \frac{ e^{z_i} }{ 1+ e^{z_i} } - \frac{1}{2} \right ) \, .
$$
Therefore, if the log prior $g$ satisfies (\ref{eqn:priormixture}), then the following three updates will generate a sequence of estimates that converges to stationary point of the posterior:
\begin{eqnarray*}
 \beta^{(g+1)} &=& \left( \tau^{-2} \Li^{(g)} + \bX_{\star}^T\Oi^{(g)} \bX_{\star} \right)^{-1} \left( \frac{1}{2} \bX_{\star}^T \bone_n \right) \, , \\
\hat\omega^{(g+1)}_i  &=& \frac{1}{z_i^{(g+1)}} \left( \frac{ e^{z_i^{(g+1)}} }{ 1+ e^{z_i^{(g+1)}} } - \frac{1}{2} \right) \, , \quad
\hat\lambda^{(g+1)}_j =  \frac{ \kappa_{\beta} + \tau^2  \ g'( \beta_j^{(g+1)} \mid \tau ) }{ \beta_j^{(g+1)} - \mu_{\beta} } \, ,
\end{eqnarray*}
where $z_i^{(g)} = y_i x_i^T \beta^{(g)}$, $\bX_{\star}$ is the matrix having rows $x^{\star}_i = y_i x_i$, and $\Omega = \mbox{diag}(\omega_1, \ldots, \omega_n)$ and $\Lambda = \mbox{diag}(\lambda_1, \ldots, \lambda_p)$ are diagonal matrices.

This sequence of steps resembles iteratively re-weighted least squares due to the presence of the diagonal weights matrix $\Omega$.  But there are subtle differences, even in the unpenalized case where $\lambda_j \equiv 0$ and the solution is the maximum-likelihood estimator.  In iteratively re-weighted least squares, the analogous weight matrix $\Omega$ has diagonal entries $\omega_i = \mu_i (1 - \mu_i)$, where $\mu_i = 1/(1+e^{-x_i^T \bbeta})$ is the estimated value of $\mbox{pr}(y_i = 1)$ at each stage of the algorithm.  These weights arise from the expected information matrix, given the current parameter estimate. They decay to zero more rapidly, as a function of the linear predictor $x_i^T \bbeta$, than the weights in our algorithm.  This can lead to numerical difficulties when the successes and failures are nearly separable by a hyperplane in $\mathbb{R}^p$, or when the algorithm is initialized far from the solution.



To illustrate this point, we ran a series of numerical experiments with pure maximum-likelihood estimation as the goal.  In each case we simulated a logistic-regression problem with standard normal coefficients $\beta_j$.  Two different design matrices were used.  The first problem had $p=100$ and $n=10^4$, and exhibited modest collinearity: each row $x_i$ was drawn from a 10-dimensional factor model, $x_i = B f_i + a_i$.  The $100 \times 10$ factor-loadings matrix $B$, the factors $f_i$, and the idiosyncratic $a_{ij}$ were all simulated from a standard normal distribution.  The second problem was larger, but nearly orthogonal: $p=500$, $n=5 \times10^4$, with each $x_{ij}$ drawn independently from a standard normal distribution.

For each data set, a logistic-regression model was estimated using iteratively re-weighted least squares; expectation--maximization, both with and without quasi-Newton acceleration; the nonlinear conjugate gradient method; and the nonlinear quasi-Newton method due to Broyden, Fletcher, Goldfarb, and Shanno.  These last two methods require the gradient of the logistic-regression likelihood, which is available in closed form.  Further details can be found in Chapters 5 and 6 of \citet{nocedal:wright:2000}.

We ran the algorithms from two different starting values: $\beta_j^{(0)} = 10^{-3}$ for all $j$, and a random starting location in the hypercube $[-1,1]^p$.  In the latter case the same random location was used for all methods.  All calculations were performed in R \citep{rcore:2012} on a standard desktop computer with 8 processor cores running at 2$\cdot$66 gigahertz.  We avoided the potential ineffiencies of R as much as possible by calling pre-compiled routines for multi-threaded matrix operations, non-linear gradient-based optimization, and iteratively re-weighted least squares. Code implementing all the experiments is available from the authors.

\begin{table}
\begin{center}
\begin{footnotesize}
\caption{\label{tab:timingresults} Run times in seconds  on two simulated problems for each logistic-regression algorithm.}
\vspace{0.5\baselineskip}
\begin{tabular}{r r r c r r}
Problem & \multicolumn{2}{c}{$p=100$, $n=10^4$} & & \multicolumn{2}{c}{$p=500$, $n=5 \times 10^4$} \\
Initial value & $\beta_j^{(0)} = 10^{-3}$ & Random & & $\beta_j^{(0)} = 10^{-3}$ & Random \\
EM &  25$\cdot$9 & 24$\cdot$8 & & 334$\cdot$8 & 321$\cdot$7 \\
Accelerated EM &  0$\cdot$8 & 1$\cdot$3 & & 13$\cdot$1  & 22$\cdot$0 \\
Quasi-Newton BFGS   & 2$\cdot$3 & 2$\cdot$3 & & 90$\cdot$3  & 88$\cdot$0 \\
Nonlinear conjugate gradient   & 1$\cdot$8 & 1$\cdot$8 & & 198$\cdot$7  & 62$\cdot$1  \\
	IRLS & 1$\cdot$9 & diverged 		& & 158$\cdot$7  & diverged
\end{tabular}
\end{footnotesize}
\end{center}
\end{table}

Table \ref{tab:timingresults} shows the run times for each method.  These timings depend upon many factors specific to a particular computing environment, including the degree to which the sub-routines of each algorithm are able to exploit a multi-core architecture, and the way optimization and matrix operations are performed in R. Thus one should not read too much into the precise values.  

Nonetheless, some tentative conclusions may be drawn.  In both the collinear and nearly orthogonal cases, the iteratively re-weighted least-squares algorithm proved sensitive to the choice of starting value.  It converged when all components of $\bbeta$ were initialized to $10^{-3}$, but not when initialized to a random point in $[-1,1]^p$.  This reflects the fact that a quadratic approximation to the logistic likelihood can be poor in regions far from the solution.  This may lead to an ill-conditioned linear system in the initial iterations of the algorithm, which is sometimes severe enough to cause divergence unless some form of trust-region strategy is used.

The basic version of the expectation-maximization algorithm is slow but robust, converging in all cases.  Moreoever, combining expectation-maximization with quasi-Newton acceleration led to an algorithm that was equally robust, and faster than all other algorithms we tried.

\subsection{Penalized logistic regression}

\begin{figure}
\begin{center}
\includegraphics[width=5.5in]{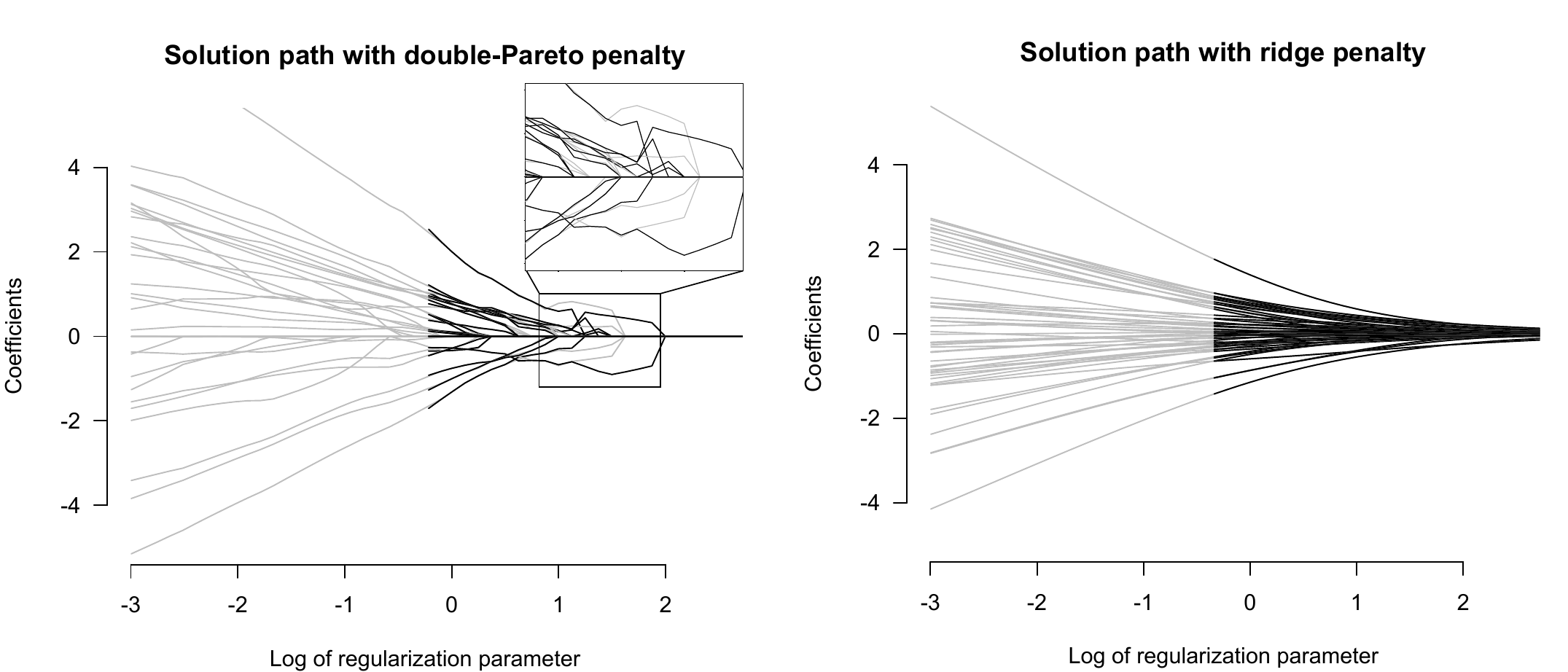}
\end{center}
\caption{\label{fig:ridgelogitpath}The solution paths for $\bbeta$ for the double-Pareto and ridge penalties, as a function of the regularization parameter $\log(1/\tau)$, for a simulated logistic regression problem.  The black lines show the solution for iteratively re-weighted penalized least squares; the grey lines, for expectation-maximization.  The black lines stop where iteratively re-weighted least squares fails due to numerical instability.}
\end{figure}

We also investigated the performance of data augmentation versus iteratively re-weighted penalized least squares.  For this case we simulated data with a nontrivial correlation structure in the design matrix.  Let $\Sigma = BB^T   +  \Psi$, where $B$ is a $50 \times 4$ matrix of standard normal random entries, and $\Psi$ is a diagonal matrix with $\chi^2_1$ random entries.  The rows of the design matrix $X$ were simulated from a multivariate normal distribution with mean zero and covariance matrix $\Sigma$, and the coefficients $\beta_j$ were standard normal random draws.  The size of the data set was $p=50$ and $n=200$.  We used a ridge-regression penalty, along with the generalized double-Pareto model where $p(\beta_j \mid \tau) \propto \{1 + |\beta_j|/(a \tau) \}^{-(1 + a)}$ \citep{dunson:armagan:lee:2010}.  This is non-differentiable at zero and is therefore sparsity-inducing, but has polynomial tails. It also has a conditionally Gaussian representation, making Proposition \ref{thm:reptheorem1} applicable.

We chose $a=2$, and used each algorithm to compute a full solution path for $\bbeta$ as a function of the regularization parameter, here expressed as $\log(1/\tau)$ in keeping with the penalized-likelihood literature.  Each solution was initially computed for $\tau_1 = 10^{-3}$, thereby constraining all coefficients to be zero or very small.  The value of $\tau$ was then increased along a discrete grid $\{\tau_1, \ldots, \tau_K=1000\}$, using the solution for $\tau_k$ as the starting value for the $\tau_{k+1}$ case.

As Figure \ref{fig:ridgelogitpath} shows, iteratively re-weighted least squares failed when $\log(1/\tau)$ became too small, causing the linear system that must be solved at each stage of the algorithm to be numerically singular.  This happened before all coefficients had entered the model, and when 20 out of 200 observations still had fitted success probabilities between 0$\cdot$05 and 0$\cdot$95.

Sparse logistic regression via penalized likelihood is a topic of great current interest \citep{genkin:lewis:madigan:2007,meier:vandegeer:buhlmann:2008}.  This problem involves three distinct issues: how to handle the logistic likelihood; how to choose a penalty function; and how to fit the resulting model.  These issues interact in poorly understood ways.  For example, coordinate-wise algorithms, including Gibbs sampling, can fare poorly in multimodal situations.  Nonconvex penalties lead to multimodal objective functions, but also, subject to certain regularity conditions, exhibit more favorable statistical properties for estimating sparse signals \citep{fan:li:2001,Carvalho:Polson:Scott:2008a}.  Moreover, coordinate descent is tractable only if the chosen penalty leads to a univariate thresholding function whose solution is analytically available \citep{mazumder:friedman:hastie:2009}.  This is a fairly narrow class, and does not include most of the penalties mentioned in the introduction.

The question of how to handle the likelihood further complicates matters.  For example, the area of detail in Figure \ref{fig:ridgelogitpath} shows that, for a double-Pareto penalty, the solution paths fit by iteratively re-weighted penalized least squares differ in subtle but noticeable ways from those fit by expectation-maximization.  By checking the maximized value of the objective function under both methods, we are able to confirm that iteratively re-weighted penalized least squares does not yield the true optimum.  Yet we do not entirely understand why, and under what circumstances, the methods will differ, and how these differences should affect recommendations about what penalty function and algorithm should be used to fit logistic regression models.  A full study of these issues is beyond the scope of the current paper, but is a subject of active inquiry.

%
%
%

\subsection{Penalized quantile regression}

Next, we show how our data-augmentation scheme can be used to fit penalized quantile-regression models, and we compare these fits to the corresponding unpenalized estimates.  Choose $p(\omega_i)$ to be a generalized inverse-Gaussian prior of unit scale, where $(\alpha, \kappa, \mu) = (1, 1- 2q, 0)$.  This gives $-\log p(z_i) = |z_i| + (2q-1) z_i$, the pseudo-likelihood which yields quantile regression for the $q$th quantile \citep{koenker:2005,li:xi:lin:2010}.  Applying Proposition \ref{thm:reptheorem1}, we find that $\hat\omega_i = |y_i - x_i^T \beta^{(g)}|^{-1}$ as the expected value of the conditional sufficient statistic needed in the expectation step of our algorithm.

To study the method, we simulated 50 data sets with $p=25$, $n=50$, and $\beta = (5,4,3,2,1,0, \ldots, 0)^T$.  In each case the 90th percentile of the data was a linear function of $\beta$ with i.i.d.~$\N(0,1)$ design points.  Noise was added by simulating errors from a normal distribution whose 90th percentile was the linear predictor $x_i^T \beta$, and whose variance was $\sigma^2 = 5^2$.  For each data set, we fit three models: traditional quantile regression using the R package from \citet{koenker:2011}, along with quantile regression under the lasso penalty and the generalized double-Pareto penalty with $\alpha = 3$.  For the second and third method, the scale of regularization $\tau$ was chosen by cross validation.  Performance was measured by squared-error loss in estimating $\beta$, and out-of-sample check loss on a new data set of the same size, simulated from the same model.

The results are in Table \ref{tab:QRimulation}.  Both regularized versions of quantile regression for the 90th percentile seem to outperform the straight estimator.  No significant differences in performance emerged between the lasso and double-Pareto penalties, although the double-Pareto solutions were systematically more sparse.

\begin{table}
\begin{center}
\caption{\label{tab:QRimulation}Results of the quantile-regression simulation study.}
\vspace{1pc}
\begin{tabular}{l r r r}
& Unpenalized & Lasso & Double Pareto \\
Estimation error & 17 & 10 & 10 \\
Out-of-sample check loss & 764 & 704 & 692 \\
Average model size & 25$\cdot$0 & 18$\cdot$2 & 13$\cdot$1
\end{tabular}
\end{center}
\end{table}

%

\section{Discussion}

Our primary goal in this paper has been to show the relevance of the conditionally Gaussian representation of (\ref{eqn:likelihoodmixture})--(\ref{eqn:priormixture}), together with Proposition \ref{thm:reptheorem1}, for fitting a wide class of regularized estimators within a unified variance-mean mixture framework.  We have therefore focused only on the most basic implementation of the expectation-maximization algorithm, together with quasi-Newton acceleration.
 
There are many variants of the expectation-maximization algorithm, however, some of which can lead to dramatic improvements \citep{meng:vandyk:1997,gelman:vandyk:huang:boscardin:2008}.  These variants include parameter expansion \citep{liu:wu:1999}, majorization--minimization \citep{hunter:lange:2000}, the partially collapsed Gibbs sampler \citep{vandyk:park:2008}, and simulation-based alternatives \citep{vandyk:meng:2011}.  Many of these modifications require additional analytical work for particular choices of $g$ and $f$.  One example here includes the work of \citet{gelman:meng:liu:2005} on the robit model.  We have not explored these options here, and this remains a promising area for future research.


A second important fact is that, for many purposes, such as estimating $\beta$ under squared-error loss, the relevant quantity of interest is the posterior mean rather than the mode.  Indeed, both \citet{hans:2008} and \citet{efron:2009} argue that, for predicting future observables, the posterior mean of $\beta$ is the better choice.  The following proposition represents the posterior mean for $\beta$ in terms of the score function of the predictive distribution, generalizing the results of \citet{brown:1971}, \citet{masreliez:1975}, \citet{pericchi:smith:1992}, and \citet{Carvalho:Polson:Scott:2008a}.  There are a number of possible versions of such a result.  Here we consider a variance-mean mixture prior $p(\beta_j)$ with a location likelihood $p( y- \beta) $, but a similar result holds the other way around.

\begin{proposition}
\label{thm:reptheorem3}
Let $p(|y - \beta_j|)$ be the likelihood for a location parameter $\beta_j$, symmetric in $y - \beta$, and let $ p( \beta_j ) = \int \phi ( \beta_j ; \mu_{\beta} + \kappa_{\beta} / \lambda_j , \tau^2  / \lambda_j ) \ p( \lambda_j^{-1} ) \ d \lambda_j^{-1} $ be a normal variance-mean mixture prior.  Define the following quantities:
\begin{eqnarray*}
m( y) &=& \int p( y - \beta_j ) p( \beta_j ) \ d \beta_j \; , \quad \quad p^\star ( \lambda_j^{-1} ) = \frac{ \lambda_j p( \lambda_j^{-1} )}{ E ( \lambda_j ) } \, , \\
p^\star ( \beta_j ) &=& \int \phi ( \beta_j ; \mu + \kappa / \lambda_j , \tau^2 / \lambda_j ) p^{\star}( \lambda_j^{-1} ) \; , \quad \quad
m^\star (y) = \int p( y - \beta_j ) p^{\star}( \beta_j ) \, .
\end{eqnarray*}
Then
$$
E( \beta_j  \mid  y) =   - \frac{\kappa_\beta}{\tau^2 } + \left\{ \frac{ \mu_\beta  E ( \lambda_j ) }{\tau^2 } \right\} \left\{ \frac{m^\star (y)}{m(y)} \right\}
 + \left\{  \frac{ E ( \lambda_j) }{ \tau^2 }\right\} \left\{  \frac{m^\star (y)}{m(y)} \right\}  \left\{  \frac{ \partial \log m^\star(y) }{\partial y} \right\} \, .
$$
\end{proposition}

The generalization to nonorthogonal designs is straightforward, following the original \citet{masreliez:1975} paper.  See, for example, \citet{griffin:brown:2010}, along with the discussion of the Tweedie formula by \citet{efron:2011}.

Computing the posterior mean will typically require sampling from the full joint posterior distribution over all parameters.  Our data-augmentation approach can lead to Markov-chain Monte Carlo sampling schemes for just this purpose \citep[e.g.][]{gelman:etal:2008}.  The key step is the identification of the conditional posterior distributions for $\lambda_j$ and $\omega_i$.  We have made some initial progress for logistic and negative-binomial models, described in a technical report available from the second author's website.  This remains an active area of research.

%

\section*{Acknowledgements}

The authors wish to thank two anonymous referees, the editor, and the associate editor for their many helpful comments in improving the paper.

\appendix

\section{Appendix A: distributional results}

\label{app:mixturetheory}

\subsection{Generalized hyperbolic distributions}

In all of the following cases, we assume that $ ( \theta \mid v ) \sim \mathcal{N} \left (  \mu + \kappa v , v \right )$, and that $v \sim p( v ) $.  Let $p(v \mid \psi, \gamma , \delta)$ be a generalized inverse-Gaussian distribution, following the notation of \citet{barndorff-nielsen:shephard:2001}.  Consider the special case where $\psi = 1$ and $\delta=0$, in which case $p(\theta)$ is a hyperbolic distribution having density
$$
p( \theta \mid \mu , \alpha , \kappa ) =
 \left( \frac{\alpha^2 - \kappa^2}{2 \alpha} \right) \exp \left\{ -\alpha | \theta - \mu | + \kappa ( \theta - \mu ) \right\}  \, .
$$
When viewed as a pseudo-likelihood or pseudo-prior, the class of generalized hyperbolic distributions will generate many common objective functions.  Choosing $(\alpha, \kappa, \mu) = (1,0,0)$ leads to $-\log p(\beta_j) = |\beta_j|$, and thus $\ell^1$ regularization.  Choosing $(\alpha, \kappa, \mu) = (1, 1- 2q, 0)$ gives $-\log p(z_i) = |z_i| + (2q-1) z_i$.  This is the check-loss function, yielding quantile regression for the $q$th quantile.  Finally, choosing $(\alpha, \kappa, \mu) = (1,1,1)$ leads to the maximum operator:
$$
-(1/2) \log p(z_i) = (1/2) |1-z_i| + (1/2)(1-z_i) = \max (1-z_i, 0) \, ,
$$
where $z_i = y_i x_i^T \beta$.  This is the objective function for support vector machines \citep[e.g.][]{mallick:ghosh:ghosh:2005,polson:stevescott:2011}, and corresponds to the limiting case of a generalized inverse-Gaussian prior.

\subsection{$Z$ distributions}

Let $p_{\pol}(v \mid \alpha, \alpha - 2\kappa)$ be a Polya distribution, which can be represented as an infinite convolution of exponentials, and leads to a $Z$ distributed marginal  \citep{bn:kent:sorensen:1982}.  The important result is the following:
$$
p_{Z} ( \theta \mid \mu , \alpha , \kappa ) = \frac{1}{B( \alpha , \kappa )} 
  \frac{ ( e^{\theta-\mu})^{\alpha} }{ ( 1 + e^{\theta-\mu} )^{2( \alpha - \kappa ) } } 
  =
\int_0^\infty \mathcal{N} \left (  \mu + \kappa v , v \right ) 
 p_{\pol} ( v \mid \alpha, \alpha - 2\kappa) \ d v \, .
$$

For logistic regression, choose $(\alpha, \kappa, \mu) = (1,1/2,0)$, leading to $p(z_i) = e^{z_i}/(1+e^{z_i})$ with $z_i = y_i x_i^T \beta$.  This corresponds to a limiting improper case of the Polya distribution.  The necessary mixture representation still holds, however, by applying the Fatou-Lebesgue theorem \citep{gramacy:polson:2012}.

For the multinomial generalization of the logistic model, we require a slight modification.  Suppose that $ y_i \in \{ 1 , \ldots , K \} $ is a category indicator, and that $\beta_k = (\beta_{k1}, \ldots, \beta_{kp})^T$ is the block of $p$ coefficients for the $k$th category. Let $\eta_{ik} = \exp \left ( x_i^T \beta_k - c_{ik} \right ) / \{1 + \exp \left ( x_i^T \beta_k - c_{ik} \right ) \}$, where $c_{ik} ( \beta_{-k} ) = \log \Big\{ \sum_{l \neq k} \exp(x_i^T \beta_l) \Big\}$.  We follow \citet{holmes:held:2006}, writing the conditional likelihood for $\beta_k$ as
\begin{align*}
L( \beta_k \mid \beta_{-k} , y ) \propto \prod_{i=1}^n \prod_{l=1}^{K} \eta_{il}^{\mathbb{I} (y_i=l)} &\propto \quad \prod_{i=1}^n \eta_{ik}^{ \mathbb{I} ( y_i = k )} \{ w_i ( 1 - \eta_{ik} ) \}^{ \mathbb{I} ( y_i \neq k )} \\
& \propto \prod_{i=1}^n
\left\{
 \frac{\exp \left ( x_i^T \beta_k - c_{ik} \right ) ^{ \mathbb{I} ( y_i = k )} }
{1+ \exp \left ( x_i^T \beta_k - c_{ik} \right )}
\right\} \, ,
\end{align*}
where $ w_i $ is independent of $\beta_k$ and $\mathbb{I}$ is the indicator function.  Thus the conditional likelihood for the $k$th block of coefficients $\beta_k$, given all the other blocks of coefficients $\beta_{-k}$, can be written as a product of $n$ terms, the $i$th term having a Polya mixture representation with $ \kappa_{ik}  = \mathbb{I}(y_i = k) - 1/2 $ and $ \mu_{ik} = c_{ik} ( \beta_{-k} ) $.  This allows regularized multinomial logistic models to be fit using the same approach of Section 4.1, with each block $\beta_k$ updated in a conditional maximization step.

\section{Appendix B: Proofs}

\subsection{Proposition \ref{thm:reptheorem1}}

\label{app:reptheorem1}

Since $ \phi $ is a normal kernel,
\begin{equation}
\label{eqn:ddxnormal}
 \frac{\partial \phi (
\beta_j \mid \mu_\beta + \kappa_\beta / \lambda_j , \tau^2/\lambda_j ) }{ \partial \beta_j} =
- \left( \frac{\beta_j -\mu_\beta - \kappa_\beta / \lambda_j}{\tau^2/\lambda_j} \right) \phi ( \beta_j \mid \mu_\beta + \kappa_\beta / \lambda_j  ,
\tau^2/\lambda_j ). 
\end{equation}

We use this fact to differentiate
$$
 p( \beta_j \mid \tau ) = \int_0^\infty \phi ( \beta_j \mid \mu_{\beta} + \kappa_\beta / \lambda_j , \tau^2/\lambda_j ) \
p( \lambda_j \mid \tau ) \ d \lambda_j
$$
under the integral sign to obtain
$$
\frac{\partial}{\partial \beta_j} p( \beta_j \mid \tau ) = \int_0^\infty \frac{\partial}{\partial \beta_j} \left\{ 
 \phi ( \beta_j \mid \mu_{\beta} + \kappa_{\beta} / \lambda_j , \tau^2 / \lambda_j )  \right\}
\ p( \lambda_j \mid \tau ) \ d \lambda_j \, .
$$

Dividing by $ p( \beta_j \mid  \tau ) $ and using  (\ref{eqn:ddxnormal}) for the inner function, we get
$$
\frac{\partial}{\partial \beta_j} p( \beta_j \mid  \tau ) = \left( \frac{\kappa_{\beta}}{\tau^2} \right) \ p( \beta_j \mid  \tau )
 - \left( \frac{\beta_j - \mu_{\beta}}{\tau^2} \right)
E\left ( \lambda_j \mid  \beta_j , \tau \right ) \, .
$$
Thus
$$
\left\{ \frac{1}{p( \beta_j \mid \tau )} \right\} \frac{\partial}{\partial \beta_j} p( \beta_j \mid \tau ) = 
\frac{\kappa_\beta}{\tau^2} - \left( \frac{\beta_j - \mu_{\beta}}{\tau^2} \right)
 E \left ( \lambda_j \mid \beta^{(g)} , \tau , y \right ) \, .
$$
Equivalently, in terms of the penalty function $-\log p(\beta_j \mid \tau)$,
$$
( \beta_j - \mu_{\beta} ) E \left ( \lambda_j \mid \beta_j   \right ) =  \kappa_\beta -
 \tau^2 \frac{\partial}{\partial \beta_j} \log p( \beta_j \mid \tau ) \, .
$$
By a similar argument,
$$
( z_i - \mu_z ) E \left ( \omega_i \mid \beta , z_i, \sigma \right ) =  \kappa_z -
 \sigma^2 \frac{\partial}{\partial z_i} \log p( z_i \mid \beta, \sigma) \, .
$$
We obtain the result using the identities
$$
\frac{\partial}{\partial z_i} \log p( z_i \mid \beta_i ) = -f^\prime ( z_i \mid \beta, \sigma)
\quad {\rm and} \quad \frac{\partial}{\partial \beta_j} \log p( \beta_j \mid \tau ) = -g^\prime ( \beta_j \mid  \tau ) \, .
$$

\subsection{Proposition \ref{thm:reptheorem2}}
\label{app:reptheorem2}

We derive the expressions for a regression problem, with those for classification involving only small changes.  Begin with Equation (\ref{eqn:compdatalogpost}).  Collecting terms, we can represent the log posterior, up to an additive constant not involving $\bbeta$, as a sum of quadratic forms in $\bbeta$:
\begin{eqnarray*}
 \log p( \beta \mid \omega, \lambda, \tau, \sigma, z ) = &-& \frac{1}{2} \left(\{y - \mu_z \bone - \kappa_z \omega^{-1} \} - X\beta \right)^T \Omega \left(\{y - \mu_z \bone - \kappa_z \omega^{-1}\} - X\beta \right) \\
 &-& \frac{1}{2\tau^2} \left( \beta - \mu_b \bone - \kappa_\beta \lambda^{-1} \right)^T \Lambda^{-1} \left( \beta - \mu_\beta \bone - \kappa_\beta \lambda^{-1} \right) \, .
\end{eqnarray*}
recalling that $\omega^{-1}$ and $\lambda^{-1}$ are column vectors.  This is the log posterior under a normal prior $\beta \sim \N(\mu_\beta \bone + \kappa_\beta \lambda^{-1}, \tau^2 \Lambda^{-1})$ after having observed the working response $y - \mu_z \bone - \kappa_z \omega^{-1}$.  The identity $\Omega (\mu \bone + \kappa \omega^{-1}) = \mu \omega + \kappa \bone$ then gives the result.

For classification, on the other hand, let $\bX_{\star}$ be the matrix with rows $x_i^{\star} = y_i x_i$.  The kernel of the conditionally normal likelihood then becomes $(\bX_{\star} \beta - \mu_z \bone - \kappa_z \omega^{-1})^T \ \Omega \ (\bX_{\star} \beta - \mu_z \bone - \kappa_z \omega^{-1})$. 
Hence it is as if we observe the $n$-dimensional working response $ \mu_z \bone + \kappa_z \omega^{-1}$ in a regression model having design matrix $\bX_{\star}$.

\subsection{Proposition \ref{thm:reptheorem3}}

\label{app:reptheorem3}

%

Our extension of Masreliez's theorem to variance-mean mixtures follows a similar path as Proposition \ref{thm:reptheorem1}. Since $ \phi $ is a normal kernel, we may apply (\ref{eqn:ddxnormal}), giving
$$
\frac{1}{\tau^2  \lambda_j} \beta_j \phi ( \beta_j \mid  \mu_\beta + \kappa_\beta / \lambda_j , \tau^2  / \lambda_j )
 = \frac{ \mu_\beta - \kappa_\beta}{\tau^2 /  \lambda_j} - \frac{\partial \phi (
\beta_j \mid \mu_\beta + \kappa_\beta / \lambda_j , \tau^2/\lambda_j ) }{ \partial \beta_j} \, .
$$
The rest of the argument follows the standard Masreliez approach.

\end{spacing}

\begin{small}
\bibliographystyle{biometrika} 
\bibliography{masterbib}

\begin{thebibliography}{50}
\expandafter\ifx\csname natexlab\endcsname\relax\def\natexlab#1{#1}\fi

\bibitem[{Andrews \& Mallows(1974)}]{andrews:mallows:1974}
\textsc{Andrews, D.F.} \& \textsc{Mallows, C.L.} (1974).
\newblock Scale mixtures of normal distributions.
\newblock \textit{Journal of the Royal Statistical Society, Series B}
  \textbf{36}, 99--102.

\bibitem[{Armagan et~al.(2012)Armagan, Dunson \& Lee}]{dunson:armagan:lee:2010}
\textsc{Armagan, A.}, \textsc{Dunson, D.} \& \textsc{Lee, J.} (2012).
\newblock Generalized double {P}areto shrinkage.
\newblock \textit{Statistica Sinica} \textbf{to appear}.

\bibitem[{Bae \& Mallick(2004)}]{bae:mallick:2004}
\textsc{Bae, K.} \& \textsc{Mallick, B.K.} (2004).
\newblock Gene selection using a two-level hierarchical {B}ayesian model.
\newblock \textit{Bioinformatics} \textbf{20}, 3423--30.

\bibitem[{Barndorff-Nielsen et~al.(1982)Barndorff-Nielsen, Kent \&
  Sorensen}]{bn:kent:sorensen:1982}
\textsc{Barndorff-Nielsen, O.E.}, \textsc{Kent, J.} \& \textsc{Sorensen, M.}
  (1982).
\newblock Normal variance-mean mixtures and z distributions.
\newblock \textit{International Statistical Review} \textbf{50}, 145--59.

\bibitem[{Barndorff-Nielsen \&
  Shephard(2001)}]{barndorff-nielsen:shephard:2001}
\textsc{Barndorff-Nielsen, O.E.} \& \textsc{Shephard, N.} (2001).
\newblock Non-{G}aussian {O}rnstein--{U}hlenbeck-based models and some of their
  uses in financial economics (with discussion).
\newblock \textit{Journal of the Royal Statistical Society: Series B
  (Statistical Methodology)} \textbf{63}, 167--241.

\bibitem[{Brown(1971)}]{brown:1971}
\textsc{Brown, L.D.} (1971).
\newblock Admissible estimators, recurrent diffusions and insoluble boundary
  problems.
\newblock \textit{The Annals of Mathematical Statistics} \textbf{42}, 855--903.

\bibitem[{Caron \& Doucet(2008)}]{caron:doucet:2008}
\textsc{Caron, F.} \& \textsc{Doucet, A.} (2008).
\newblock Sparse {B}ayesian nonparametric regression.
\newblock In \textit{Proceedings of the 25th International Conference on
  Machine Learning}. Helsinki, Finland: Association for Computing Machinery,
  pp. 88--95.

\bibitem[{Carvalho et~al.(2010)Carvalho, Polson \&
  Scott}]{Carvalho:Polson:Scott:2008a}
\textsc{Carvalho, C.M.}, \textsc{Polson, N.G.} \& \textsc{Scott, J.G.}
  (2010).
\newblock The horseshoe estimator for sparse signals.
\newblock \textit{Biometrika} \textbf{97}, 465--80.

\bibitem[{Dempster et~al.(1977)Dempster, Laird \&
  Rubin}]{dempster:laird:rubin:1977}
\textsc{Dempster, A.P.}, \textsc{Laird, N.M.} \& \textsc{Rubin, D.B.} (1977).
\newblock Maximum likelihood from incomplete data via the {EM} algorithm (with
  discussion).
\newblock \textit{Journal of the Royal Statistical Society (Series B)}
  \textbf{39}, 1--38.

\bibitem[{Efron(2009)}]{efron:2009}
\textsc{Efron, B.} (2009).
\newblock Empirical {B}ayes estimates for large-scale prediction problems.
\newblock \textit{Journal of the American Statistical Association}
  \textbf{104}, 1015--28.

\bibitem[{Efron(2011)}]{efron:2011}
\textsc{Efron, B.} (2011).
\newblock {T}weedie's formula and selection bias.
\newblock \textit{Journal of the American Statistical Association}
  \textbf{106}, 1602--14.

\bibitem[{Fan \& Li(2001)}]{fan:li:2001}
\textsc{Fan, J.} \& \textsc{Li, R.} (2001).
\newblock Variable selection via nonconcave penalized likelihood and its oracle
  properties.
\newblock \textit{Journal of the American Statistical Association} \textbf{96},
  1348--60.

\bibitem[{Figueiredo(2003)}]{figueiredo:2003}
\textsc{Figueiredo, M.} (2003).
\newblock Adaptive sparseness for supervised learning.
\newblock \textit{IEEE Transactions on Pattern Analysis and Machine
  Intelligence} \textbf{25}, 1150--9.

\bibitem[{Gelman(2006)}]{gelman:2006}
\textsc{Gelman, A.} (2006).
\newblock Prior distributions for variance parameters in hierarchical models.
\newblock \textit{Bayesian Analysis} \textbf{1}, 515--33.

\bibitem[{Gelman et~al.(2008{\natexlab{a}})Gelman, Jakulin, Pittau \&
  Su}]{gelman:etal:2008}
\textsc{Gelman, A.}, \textsc{Jakulin, A.}, \textsc{Pittau, M.} \& \textsc{Su,
  Y.} (2008{\natexlab{a}}).
\newblock A weakly informative default prior distribution for logistic and
  other regression models.
\newblock \textit{The Annals of Applied Statistics} \textbf{2}, 1360--83.

\bibitem[{Gelman et~al.(2008{\natexlab{b}})Gelman, van Dyk, Huang \&
  Boscardin}]{gelman:vandyk:huang:boscardin:2008}
\textsc{Gelman, A.}, \textsc{van Dyk, D.A.}, \textsc{Huang, Z.} \&
  \textsc{Boscardin, W.J.} (2008{\natexlab{b}}).
\newblock Using redundant parameterizations to fit hierarchical models.
\newblock \textit{Journal of Computational and Graphical Statistics}
  \textbf{17}, 95--122.

\bibitem[{Genkin et~al.(2007)Genkin, Lewis \&
  Madigan}]{genkin:lewis:madigan:2007}
\textsc{Genkin, A.}, \textsc{Lewis, D.D.} \& \textsc{Madigan, D.} (2007).
\newblock Large-scale Bayesian logistic regression for text categorization.
\newblock \textit{Technometrics} \textbf{49}, 291--304.

\bibitem[{Gramacy \& Polson(2012)}]{gramacy:polson:2012}
\textsc{Gramacy, R.B.} \& \textsc{Polson, N.G.} (2012).
\newblock Simulation-based regularized logistic regression.
\newblock \textit{Bayesian Analysis} \textbf{7}, 567--90.

\bibitem[{Griffin \& Brown(2010)}]{griffin:brown:2010}
\textsc{Griffin, J.E.} \& \textsc{Brown, P.J.} (2010).
\newblock Inference with normal-gamma prior distributions in regression
  problems.
\newblock \textit{Bayesian Analysis} \textbf{5}, 171--88.

\bibitem[{Griffin \& Brown(2012)}]{griffin:brown:2005}
\textsc{Griffin, J.E.} \& \textsc{Brown, P.J.} (2012).
\newblock Alternative prior distributions for variable selection with very many
  more variables than observations.
\newblock \textit{Australian and New Zealand Journal of Statistics} (to
  appear).

\bibitem[{Hans(2009)}]{hans:2008}
\textsc{Hans, C.M.} (2009).
\newblock {B}ayesian lasso regression.
\newblock \textit{Biometrika} \textbf{96}, 835--45.

\bibitem[{Holmes \& Held(2006)}]{holmes:held:2006}
\textsc{Holmes, C.C.} \& \textsc{Held, L.} (2006).
\newblock Bayesian auxiliary variable models for binary and multinomial
  regression.
\newblock \textit{Bayesian Analysis} \textbf{1}, 145--68.

\bibitem[{Huang et~al.(2008)Huang, Horowitz \& Ma}]{huang:horowitz:ma:2008}
\textsc{Huang, J.}, \textsc{Horowitz, J.L.} \& \textsc{Ma, S.} (2008).
\newblock Asymptotic properties of bridge estimators in sparse high-dimensional
  regression models.
\newblock \textit{The Annals of Statistics} \textbf{36}, 587--613.

\bibitem[{Hunter \& Lange(2000)}]{hunter:lange:2000}
\textsc{Hunter, D.R.} \& \textsc{Lange, K.} (2000).
\newblock Quantile regression via an {MM} algorithm.
\newblock \textit{Journal of Computational and Graphical Statistics}
  \textbf{9}, 60--77.

\bibitem[{Knight \& Fu(2000)}]{knight:fu:1998}
\textsc{Knight, K.} \& \textsc{Fu, W.} (2000).
\newblock Asymptotics for lasso-type estimators.
\newblock \textit{The Annals of Statistics} \textbf{28}, 1356--78.

\bibitem[{Koenker(2005)}]{koenker:2005}
\textsc{Koenker, R.} (2005).
\newblock \textit{Quantile {R}egression}.
\newblock New York, USA: Cambridge University Press.

\bibitem[{Koenker(2011)}]{koenker:2011}
\textsc{Koenker, R.} (2011).
\newblock \textit{quantreg: Quantile Regression}.
\newblock R package version 4.76.

\bibitem[{Koenker \& Machado(1999)}]{koenker:machado:1999}
\textsc{Koenker, R.} \& \textsc{Machado, J.} (1999).
\newblock Goodness of fit and related inference processes for quantile
  regression.
\newblock \textit{Journal of the American Statistical Association} \textbf{94},
  1296--1310.

\bibitem[{Lange(1995)}]{lange:1995}
\textsc{Lange, K.} (1995).
\newblock A quasi-{N}ewton acceleration of the {EM} algorithm.
\newblock \textit{Statistica Sinica} \textbf{5}, 1--18.

\bibitem[{Li et~al.(2010)Li, Xi \& Lin}]{li:xi:lin:2010}
\textsc{Li, Q.}, \textsc{Xi, R.} \& \textsc{Lin, N.} (2010).
\newblock Bayesian regularized quantile regression.
\newblock \textit{Bayesian Analysis} \textbf{5}, 533--56.

\bibitem[{Liu(1995)}]{liu:1995}
\textsc{Liu, C.} (1995).
\newblock Missing data imputation using the multivariate t distribution.
\newblock \textit{Journal of Multivariate Analysis} \textbf{53}, 139--58.

\bibitem[{Liu(2005)}]{gelman:meng:liu:2005}
\textsc{Liu, C.} (2005).
\newblock Robit regression: a simple robust alternative to logistic and probit
  regression.
\newblock In \textit{Applied Bayesian Modeling and Causal Inference from
  Incomplete-Data Perspectives: An Essential Journey with Donald Rubin's
  Statistical Family}, A.~Gelman \& X.L. Meng, eds., chap.~21. John Wiley \&
  Sons (London), pp. 227--38.

\bibitem[{Liu \& Wu(1999)}]{liu:wu:1999}
\textsc{Liu, J.S.} \& \textsc{Wu, Y.N.} (1999).
\newblock Parameter expansion for data augmentation.
\newblock \textit{Journal of the American Statistical Association} \textbf{94},
  1264--74.

\bibitem[{Mallick et~al.(2005)Mallick, Ghosh \&
  Ghosh}]{mallick:ghosh:ghosh:2005}
\textsc{Mallick, B.K.}, \textsc{Ghosh, D.} \& \textsc{Ghosh, M.} (2005).
\newblock Bayesian classification of tumours by using gene expression data.
\newblock \textit{Journal of the Royal Statistical Society (Series B)}
  \textbf{67}, 219--34.

\bibitem[{Masreliez(1975)}]{masreliez:1975}
\textsc{Masreliez, C.} (1975).
\newblock Approximate non-{G}aussian filtering with linear state and
  observation relations.
\newblock \textit{IEEE.~Trans.~Autom.~Control} \textbf{20}, 107--10.

\bibitem[{Mazumder et~al.(2011)Mazumder, Friedman \&
  Hastie}]{mazumder:friedman:hastie:2009}
\textsc{Mazumder, R.}, \textsc{Friedman, J.H.} \& \textsc{Hastie, T.} (2011).
\newblock Sparsenet: coordinate descent with non-convex penalties.
\newblock \textit{Journal of the American Statistical Association}
  \textbf{106}, 1125--38.

\bibitem[{Meier et~al.(2008)Meier, van~de Geer \&
  B\"uhlmann}]{meier:vandegeer:buhlmann:2008}
\textsc{Meier, L.}, \textsc{van~de Geer, S.} \& \textsc{B\"uhlmann, P.} (2008).
\newblock The group lasso for logistic regression.
\newblock \textit{Journal of the Royal Statistical Society (Series B)}
  \textbf{70}, 53--71.

\bibitem[{Meng \& van {D}yk(1997)}]{meng:vandyk:1997}
\textsc{Meng, X.L.} \& \textsc{van {D}yk, D.A.} (1997).
\newblock The {EM} algorithm---an old folk-song sung to a fast new tune (with
  discussion).
\newblock \textit{Journal of the Royal Statistical Society (Series B)}
  \textbf{59}, 511--67.

\bibitem[{Nocedal \& Wright(2000)}]{nocedal:wright:2000}
\textsc{Nocedal, J.} \& \textsc{Wright, S.J.} (2000).
\newblock \textit{Numerical Optimization}.
\newblock New York, NY, USA: Springer.

\bibitem[{Park \& Casella(2008)}]{park:casella:2008}
\textsc{Park, T.} \& \textsc{Casella, G.} (2008).
\newblock The {B}ayesian lasso.
\newblock \textit{Journal of the American Statistical Association}
  \textbf{103}, 681--6.

\bibitem[{Pericchi \& Smith(1992)}]{pericchi:smith:1992}
\textsc{Pericchi, L.R.} \& \textsc{Smith, A.} (1992).
\newblock Exact and approximate posterior moments for a normal location
  parameter.
\newblock \textit{Journal of the Royal Statistical Society (Series B)}
  \textbf{54}, 793--804.

\bibitem[{Polson \& Scott(2011{\natexlab{a}})}]{Polson:Scott:2010a}
\textsc{Polson, N.G.} \& \textsc{Scott, J.G.} (2011{\natexlab{a}}).
\newblock Shrink globally, act locally: sparse {B}ayesian regularization and
  prediction (with discussion).
\newblock In \textit{Proceedings of the 9th Valencia World Meeting on Bayesian
  Statistics}, J.M. Bernardo, M.J. Bayarri, J.O. Berger, A.P. Dawid,
  D.Heckerman, A.F.M. Smith \& M.West, eds. Oxford University Press, pp.
  501--38.

\bibitem[{Polson \& Scott(2012)}]{polson:scott:2009a}
\textsc{Polson, N.G.} \& \textsc{Scott, J.G.} (2012).
\newblock Good, great, or lucky? {S}creening for firms with sustained superior
  performance using heavy-tailed priors.
\newblock \textit{The Annals of Applied Statistics} \textbf{6}, 161--85.

\bibitem[{Polson \& Scott(2011{\natexlab{b}})}]{polson:stevescott:2011}
\textsc{Polson, N.~G.} \& \textsc{Scott, S.} (2011{\natexlab{b}}).
\newblock Data augmentation for support vector machines (with discussion).
\newblock \textit{Bayesian Analysis} \textbf{6}, 1--24.

\bibitem[{{R Core Team}(2012)}]{rcore:2012}
\textsc{{R Core Team}} (2012).
\newblock \textit{{R}: A Language and Environment for Statistical Computing}.
\newblock R Foundation for Statistical Computing, Vienna, Austria.

\bibitem[{Tibshirani(1996)}]{tibshirani1996}
\textsc{Tibshirani, R.} (1996).
\newblock Regression shrinkage and selection via the lasso.
\newblock \textit{Journal of the Royal Statistical Society (Series B)}
  \textbf{58}, 267--88.

\bibitem[{Tipping(2001)}]{tipping:2001}
\textsc{Tipping, M.E.} (2001).
\newblock Sparse {B}ayesian learning and the relevance vector machine.
\newblock \textit{Journal of Machine Learning Research} \textbf{1}, 211--44.

\bibitem[{van {D}yk \& Meng(2010)}]{vandyk:meng:2011}
\textsc{van {D}yk, D.A.} \& \textsc{Meng, X.L.} (2010).
\newblock Cross-fertilizing strategies for better {EM} mountain climbing and
  {DA} field exploration: A graphical guide book.
\newblock \textit{Statistical Science} \textbf{25}, 429--49.

\bibitem[{van {D}yk \& Park(2008)}]{vandyk:park:2008}
\textsc{van {D}yk, D.A.} \& \textsc{Park, T.} (2008).
\newblock Partially collapsed {G}ibbs samplers: theory and methods.
\newblock \textit{Journal of American Statistical Association} \textbf{103},
  790--6.

\bibitem[{West(1987)}]{west:1987}
\textsc{West, M.} (1987).
\newblock On scale mixtures of normal distributions.
\newblock \textit{Biometrika} \textbf{74}, 646--8.

\end{thebibliography}
\end{small}

\end{document}